\documentclass{article}
\usepackage{latexsym}
\usepackage{graphicx}

\begin{document}

\title{MULTILAYER SPHERICAL STELLAR CLUSTERS WITH UNIFORM DENSITY}

\author{G. S. Bisnovatyi-Kogan\footnote{Institute of Space Studies IKI, Russian Academy of Sciences, Moscow;
National Research Nuclear University MIFI, Moscow; Moscow Physical-Technical
Institute MFTI, Dolgoprudnyi;
Email: gkogan@iki.rssi.ru}}

\date{}

 \maketitle

\begin{abstract}
{Various solutions of the kinetic equation for the equilibrium of a gravitating sphere of uniform density with a quadratic gravitational potential and a linear dependence of gravitational force on radius are examined. New analytic solutions are obtained for a uniform sphere with a hollow spherical volume and central mass inside the sphere. Solutions are also obtained for an arbitrary number of spherical layers with the same density, but with different equilibrium distribution functions.}
\end{abstract}


\section{Introduction}

Because of their simplicity, uniform clusters with a quadratic gravitational
potential have been examined analytically by many authors [1-7]. Solutions
have been obtained for uniform cylindrical models with an elliptical cross
section and ellipsoidal figures of uniform density, as well as elliptical
plane disks with a surface density distribution $\sigma(x,y)$ of the form

\begin{equation}
\label{eq1}
\sigma(x,y)=\sigma_0\sqrt{1-\frac{x^2}{\alpha^2}-\frac{y^2}{\beta^2}}
\end{equation}
The equilibrium of collisionless, self-gravitating clusters is determined by
a solution that follows from the collisionless Boltzmann kinetic equation
with a self-consistent gravitational field determined by the Poisson
equation. This approach was first formulated by J. Jeans for studies of
stellar clusters [8,9] and was subsequently used by A. A. Vlasov to study
collisionless plasmas in electric and magnetic fields [10]. Analytic
equilibrium solutions for circular and elliptical disks were obtained in
Refs. 3, 11, and 12. Solutions were obtained for equilibrium cylinders with
circular and elliptical cross sections in Refs. 1, 12, and 13. Solutions
were obtained in Refs. 2 and 12 for uniform ellipsoids and spheroids
(ellipsoids of rotation). In the latter case the solution was obtained for
the case where gravitational attraction in the directions perpendicular to
the axis of symmetry of the spheroid is balanced by a centrifugal force and
the resulting solution for a uniform sphere is not spherically symmetric in
velocity space. The first equilibrium model of a cluster consisting of stars
rotating in circular orbits that was spherically symmetric in velocity space
was constructed by Einstein [14] in the framework of the general theory of
relativity for an arbitrary spherically symmetric density distribution,
including a uniform one. The stability of a Newtonian analog of the uniform
model of circular gravitating particles was studied in Refs. 15 and 16. A
uniform model, that was spherically symmetric in velocity space, was
constructed in Ref. 17, while a model with radial orbits was examined in
Ref. 18. In this paper solutions are obtained for a uniform sphere with
noncircular particle trajectories that conserve spherical symmetry in a
phase space with a hollow sphere located at its center. New solutions are
also obtained which consist of an arbitrary number of spherical layers with
equal density but with different equilibrium distribution functions.

\section{Integrals of motion and particle trajectories}

The kinetic equation for the distribution function
$f(r$,$v_{r}$,$v_{t})$ of stars in a stationary, spherically symmetric cluster with
constant density $\rho$ in a spherical coordinate system ($r,\theta,\phi,v_r,v_\theta,v_\phi$),
is written in the form [19]

\begin{eqnarray}
\label{eq2}
v_r\frac{\partial f}{\partial r}+\left(\frac{v_t^2}{r}-\frac{Gm}{r^2}\right)\frac{\partial f}{\partial v_r}-\frac{v_r v_t}{r}\frac{\partial f}{\partial v_t}=0,\nonumber\\
2\pi \int\int f dv_r v_t dv_t = \frac{\rho}{m},\quad
m=\frac{4\pi}{3}\rho r^3.
\end{eqnarray}
Here the integral is taken over a limited region of phase space and the
following conditions are used:

\begin{equation}
\label{eq3}
\frac{\partial }{\partial \theta}=\frac{\partial }{\partial \phi}=0,\quad v_t^2=v_\theta^2+v_\phi^2.
\end{equation}
The solution of Eq. (2) is a function of two integrals of motion, the
angular momentum $L$ and the total energy $E$:

\begin{eqnarray}
\label{eq4}
L=rv_t,\quad E=\frac{v_r^2}{2}+\frac{L^2}{2r^2}+\frac{2\pi}{3} G\rho  r^2-\frac{3GM}{2r_0}\nonumber\\
=\frac{v_r^2}{2}+\frac{v_t^2}{2}+\frac{2\pi}{3} G\rho  r^2-\frac{3GM}{2r_0},
\quad M=\frac{4\pi}{3}\rho r_0^3.
\end{eqnarray}
Here the notation $M,$ $\rho$, and $r_{0},$ respectively, denote the mass, density, and
radius of the spherical cluster, of which two are specified arbitrarily. In
determining the energy E the arbitrary constant is chosen from the
continuity condition for the energy at the boundary with the vacuum, where
only the gravitational energy $E_{g} =$\textit{ -GM/r} remains, that goes to zero at
infinity, and equals $E_{g0}=-\frac{GM}{r_0}$
at the boundary of the
cluster. The radial dependences of the stellar velocities for different
values of $E$ and $L$ are given by

\begin{eqnarray}
\label{eq5}
v_r^2=2\left(E+\frac{3GM}{2r_0}-\frac{L^2}{2 r^2}-\frac{2\pi}{3} G\rho  r^2\right)=2\left(E^{'}-\frac{L^2}{2 r^2}-\frac{2\pi}{3} G\rho  r^2\right),\nonumber\\
\quad v_t=\frac{L}{r},\quad E^{'}=E+\frac{3GM}{2r_0}.\qquad\qquad
\end{eqnarray}
The solution for a particle motion trajectory in a gravitational potential
$\Phi=\frac{kr^2}{2}$
has been found in Ref. 20. The closed trajectories of the motion
are ellipses with a center at the coordinate origin. The maximum and minimum
values of the radius of the trajectory \textit{r\textpm } are reached for $v_{r}=0$ and
are found from Eq. (5) in the form

\begin{equation}
\label{eq6}
r^2_{\pm}=\frac{3 E^{'}}{4\pi  G\rho } \pm \sqrt{\left(\frac{3 E^{'}}{4\pi
G\rho }\right)^2-\frac{3L^2}{4\pi G\rho }}
\end{equation}
The radius of maximum separation of a particle from the center $r_{+}$, equal
to the major semiaxis of the ellipse of the trajectory corresponds to zero
radial velocity $v_{r}$ and the minimum tangential velocity $v_{t}$, so that

\begin{equation}
\label{eq7}
r^2_{+}=\frac{3 E^{'}}{4\pi  G\rho } + \sqrt{\left(\frac{3 E^{'}}{4\pi
G\rho }\right)^2-\frac{3L^2}{4\pi G\rho }}, \quad v_{r+}=0,\quad
v_{t+}= \frac{L}{r_+}.
\end{equation}
Thus, for the minimum radius of the particle separation from the center
equal to the minor semiaxis of the ellipse of the trajectory, the radial
velocity passes through zero and the tangential velocity, through its
maximum,

\begin{equation}
\label{eq8}
r^2_{-}=\frac{3 E^{'}}{4\pi G \rho} - \sqrt{\left(\frac{3 E^{'}}{4\pi G \rho}
 \right)^2-\frac{3L^2}{4\pi G\rho}}, \quad v_{r-}=0,\quad
v_{t-}= \frac{L}{r_-}.
\end{equation}
The maximum radial velocity $v_{rm}$ is reached at radius $r_{vrm}$, when the
derivative $\frac{d(v_r^2)}{dr}$ equals zero. From Eq. (5) we have

\begin{equation}
\label{eq8a}
r_{vrm}^2=\left(\frac{3L^2}{4\pi G\rho}\right)^{1/2}, \quad
v_{rm}^2=2\left[E^{'}-\left(\frac{4\pi G \rho}{3}L^2\right)^{1/2}\right].
\end{equation}
When the expression under the root in Eqs. (6)-(8) equals zero, we obtain
the equality $r_{+}=r_-$ , which determines a circular
orbit, for which

 \begin{equation}
\label{eq9}
r^2=r_+^2=r^2_{-}=\frac{3 E^{'}}{4\pi G \rho}, \quad v_{r-}=0,\quad
v_*=v_{t+}=v_{t-}= \frac{L}{r}, \quad E^{'}
=L \sqrt{\frac{4\pi G\rho}{3}}.
\end{equation}
With Eq. (4), this yields the velocity of the motion along a circle

 \begin{equation}
\label{eq10}
v_*=\pm r\sqrt{\frac{4\pi G \rho}{3}},
\end{equation}
when the centrifugal force balances the gravitation of the sphere inside
radius r. For $L = 0$ all the particles move along radii, so that

 \begin{eqnarray}
\label{eq11}
v_t=0, \quad v_r=\sqrt {2\left( E^{'}-\frac{2\pi}{3}G \rho r^2 \right)};
\quad v_r=0\quad {\mbox{at}}\quad   r=r_0,\nonumber\\
\quad E^{'}=
\frac{2\pi}{3}G\rho r_0^2=\frac{GM}{2r_0},\quad E=-\frac{GM}{r_0}.\qquad\qquad
\end{eqnarray}

\section{Analytic spherically symmetric solutions for a sphere}

\subsection{ Uniform sphere. }

The distribution function for a uniform
Einstein sphere with radius $r$ is given by [16]

 \begin{equation}
\label{eq12}
f=\frac{\rho}{\pi m}\delta(v_r)\delta(v_t^2-\frac{4\pi G\rho}{3}r^2), \quad r\le r_0.
\end{equation}
Using the integrals of motion (4) and (5) for a sphere with radius $r_0$ the
solution can be written in the form

 \begin{equation}
\label{eq13}
f=\frac{\rho}{\pi m}\delta\left(2E^{'}\pm 2L\sqrt\frac{4\pi G\rho}{3}\right)
=\frac{\rho}{\pi m}\delta\left[v_r^2+ \left(v_t\pm \sqrt\frac{4\pi G\rho}
{3}r\right)^2\right]
\end{equation}
A distribution function of the following form for a uniform sphere that is spherically symmetric in phase space is
examined in \cite{psh73}:

 \begin{equation}
\label{eq13a}
f=\frac{\rho}{\pi^2 m \Omega^2}
\frac{\theta{\left(\frac{L^2}{2}+\frac{\Omega^2}{2}-E^{'}\right)}}{ \sqrt{{L^2}+
\Omega^2 -2E^{'}}}
=\frac{\rho}{\pi^2 m}
\frac{\theta[(1-r^2)(1-v_t^2)-v_r^2]}{\sqrt{(1-r^2)(1-v_t^2)-v_r^2}}.
\end{equation}
Here $\theta(x)$ is the Heaviside function, equal to unity for a positive and zero for a negative value of the argument. In
the latter expression (\ref{eq13a}) it is assumed $\Omega^2=\frac{4\pi G\rho}{3}=1$, and
$2E^{'}=v_r^2+v_t^2+\Omega^2 r^2=v_r^2+v_t^2+r^2$. The
equation for the density is written in the form

 \begin{equation}
\label{eq13b}
\rho=\pi m \int\int f dv_r dv_t^2= \frac{\rho}{\pi}\int\int
\frac{\theta[(1-r^2)(1-v_t^2)-v_r^2]}{\sqrt{(1-r^2)(1-v_t^2)-v_r^2}} dv_r dv_t^2.
\end{equation}
The limits of integration over phase space are determined by the condition
that the argument of the function e(x) be positive. For this condition,
after integration with respect to $dv^2$ and
$dv_r$ we obtain

\begin{eqnarray}
\label{eq13c}
\int_0^{1-\frac{v_r^2}{1-r^2}} \frac{dv_t^2}{\sqrt{(1-r^2)(1-v_t^2)-v_r^2}} =
2\frac{\sqrt{1-r^2-v_r^2}}{1-r^2},\nonumber\\ \int_{-\sqrt{1-r^2}}^{\sqrt{1-r^2}}
dv_r \frac{\sqrt{1-r^2-v_r^2}}{1-r^2}=\int_{-1}^1\sqrt{1-x^2} dx=\frac{\pi}{2}.
\end{eqnarray}
Thus, in the expression for p we obtain an identity in Eq. (15), which
confirms the self consistency of the choice of the distribution function
(15).

The distribution function for a uniform sphere with purely radial orbits can
be written in the form [18]

 \begin{equation}
\label{eq14}
f=\frac{\rho}{\pi^2 m}\frac{\delta(v_t^2)}
{\sqrt{-2E^{'}+\frac{GM}{r_0}}}= \frac{\rho}{\pi^2 m}\frac{\delta(v_t^2)}
{\sqrt{\frac{4\pi G\rho}{3}(r_0^2-r^2)-v_r^2}}.
\end{equation}

\subsection{ Sphere with a cut-out core. }If all the particles have angular
momentum $L,$ then the minimum distance of a particle from the center is
$^{r}$\textit{min }$=^{L}K$ , so an empty sphere of this radius develops inside a
spherical region of this sort. If the density of matter around this sphere
is uniform, it is easy to find analytically the trajectory of the particles
flying outside the empty sphere. The presence of an empty sphere, however,
leads to deviations from a quadratic dependence of the gravitational
potential in the material, where the trajectories of the particles are no
longer closed. An analytic solution cannot be found under these conditions.
In order to conserve the quadratic dependence and obtain an analytic
solution for an equilibrium spherically symmetric cluster, in the center of
the cluster we place a body formed by compression into a point of mass
$M_c=\frac{4\pi\rho r_{min}^3}{3}$, equal to the mass of the uniform sphere with the
radius of the void and with the density of the cluster. In this case,
gravitation in the substance surrounding the hollow sphere will be the same
as in the case of the completely uniform sphere and Eqs.(4)-(13) remain
valid for $r\ge r_{min}$

Let us consider a uniform sphere of radius $r_{0}$ with a hollow sphere of
radius $r_{\}}$ at its center. This configuration arises from a cluster of
bodies moving along identical ellipses around the center of a sphere for
which the minor and major semiaxes of the ellipse, $r_{t}$ and $r_{0}$ are
equal, respectively, to $r$ and $r+$ as defined in Eqs. (7) and (8). For
specified / and $r_{0}$ we find the integrals of motion $E'$ and $L$ of the
particles to be

  \begin{equation}
\label{eq15}
E^{'}_0=\frac{2\pi G \rho}{3}(r_0^2+r_1^2), \quad
L_0= \frac{4\pi G \rho}{3} r_0^2 r_1^2.
\end{equation}
The distribution function of the uniform sphere with a vacuum sphere at its
center (see Fig.1 \footnote{This figure was drawn by O. Yu. Tsupko, to whom the author is sincerely grateful.}) is given by

  \begin{eqnarray}
\label{eq16}
f=\frac{\rho}{\pi^2m}\frac{\delta(v_t^2-\frac{L_0^2}{r^2})}{\sqrt{2E^{'}_0-2E^{'}}}
=\frac{\rho}{\pi^2m}\frac{\delta(v_t^2-\frac{4\pi G\rho}{3}\frac{r_0^2 r_1^2}{r^2})}
{\sqrt{\frac{4\pi G \rho}{3}(r_0^2+r_1^2)-v_r^2-v_t^2- \frac{4\pi G \rho}{3} r^2}}
\nonumber\\
 =\frac{\rho}{\pi^2m}\frac{\delta(v_t^2-\frac{4\pi G\rho}{3}\frac{r_0^2 r_1^2}{r^2})}
{\sqrt{\frac{4\pi G \rho}{3}(r_0^2+r_1^2-r^2-\frac{r_0^2 r_1^2}{r^2}) -v_r^2}}.
\qquad\qquad
\end{eqnarray}
The solution (20) is a generalization of the solution (18) for purely radial
orbits to the case of particles with nonzero angular momentum: for
\textsc{L}$_{q}$\textsc{ }$=$\textsc{ }$r_{1} \quad =$ 0, Eq. (20) yields the
solution (18).

\begin{figure}[htbp]
\centerline{\includegraphics[width=1.90in,height=1.89in]{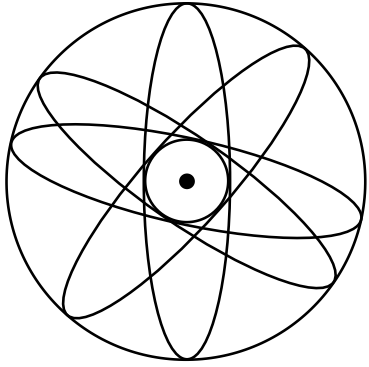}}
\caption{A schematic figure of a
sphere with a hollow core and a
central gravitational center (for
example, a supermassive black
hole) and trajectories of particles
in the form of identical ellipsoids
with centers at the coordinate
origin.}
\label{fig1}
\end{figure}

\subsection{ Multilayer sphere with uniform density.}

The solutions (18) and
(20) can be used to construct a more complicated equilibrium model for a
sphere with uniform density consisting of several layers with different
distribution functions. At the center of the sphere with a hollow core,
instead of a central gravitating mass it is possible to place a sphere with
radial orbits and to obtain a solution of the form

\begin{eqnarray}
\label{eq17}
f= \frac{\rho}{\pi^2 m}\frac{\delta(v_t^2)}
{\sqrt{\frac{4\pi G\rho}{3}(r_1^2-r^2)-v_r^2}} \quad {\mbox{for}}
\quad r\le r_1; \qquad\qquad\\
f =\frac{\rho}{\pi^2m}\frac{\delta(v_t^2-\frac{4\pi G\rho}{3}
\frac{r_1^2 r_2^2}{r^2})}{\sqrt{\frac{4\pi G \rho}{3}(r_1^2+r_2^2-r^2-
\frac{r_1^2 r_2^2}{r^2}) -v_r^2}}.  \quad {\mbox{for}} \quad
 r_1 \le r\le r_2.
\label{eq17a}
\end{eqnarray}
This sphere can be surrounded by yet another layer of matter with the
distribution function (22), in which instead of the boundaries of the layers
($r_1$, $r_{2})$ the boundaries ($r_{2}$, $r_{3})$ are specified. In the
same way it is possible to obtain a multilayer sphere model where each layer
is determined by a distribution function of the type (22) and a specified
thickness ($r_{i+1}-r_i $), where $0 \le r_{1} \le r_{2}\le...
r_{0}$ and $r_{0}$ is the outer boundary of the sphere with uniform density.

In the central region of a multilayer sphere, besides a configuration with
radial orbits (18), there may be a sphere with circular orbits (13), as well
as a sphere with a more complicated distribution function (17). Instead of
an any spherical layer at $r_{k}<r<r_{k+1}$ with the distribution
function (20), there may be a layer with circular orbits and the
distribution function (13).

\begin{enumerate}
\item \textbf{Discussion}
\end{enumerate}
The solutions of the kinetic equation for a uniform sphere of gravitating
particles (stars) have been obtained by various authors, many of whom
examined distribution functions that are not spherically symmetric [2,5,6].
Three types of models with a spherically symmetric distribution function
have also been examined: a model with circular orbits [14], a model with
radial orbits [18], and a model of a more general type [17]. In this paper
new solutions have been obtained for a uniform sphere that has a hollow
sphere with a central gravitating mass at its center which may simulate a
region of a galactic core with a supermassive black hole at its center. The
resulting solutions describe a model with elliptical trajectories that are
intermediate between radial (L$=$0) and circular ($v_{r}= 0$ )
trajectories. Based on these solutions we obtain a multilayer spherical
model with a uniform density for both hollow and filled cores, with
different distribution functions in the core and in an separate layer which
must belong to one of the four basic solutions indicated above.

This work was partially supported by RFFI grant 20-02-000455.

\end{document}